\documentclass[journal]{IEEEtran}

\usepackage{xr}
\usepackage{cite}
\usepackage{xcolor}
\usepackage{graphicx}
\usepackage{amsmath}
\usepackage{amssymb}
\usepackage{dsfont}
\usepackage{multirow}
\usepackage{soul}
\usepackage{algpseudocode}
\usepackage{algorithm}

\DeclareMathOperator*{\argmin}{arg\,min}

\usepackage{xr}
\makeatletter
\newcommand*{\addFileDependency}[1]{
  \typeout{(#1)}
  \@addtofilelist{#1}
  \IfFileExists{#1}{}{\typeout{No file #1.}}
}
\makeatother

\definecolor{new_green}{RGB}{0,120,23}

\newcommand{\newtext}[1]{\textcolor{black}{#1}}

\newenvironment{algocolor}{%
   \setlength{\parindent}{0pt}  
   \color{black}
}{}
\begin{document}

\title{Coordinate-Based Seismic Interpolation in Irregular Land Survey: A Deep Internal Learning Approach}

%
%


\author{Paul Goyes-Pe\~nafiel$^*$, 
        Edwin Vargas$^*$, ~\IEEEmembership{Student Member,~IEEE,}
        Claudia V. Correa,~\IEEEmembership{Member,~IEEE,}
        Yu Sun,~\IEEEmembership{Student Member,~IEEE,}
        Ulugbek~S.~Kamilov,~\IEEEmembership{Senior Member,~IEEE,}
        Brendt Wohlberg,~\IEEEmembership{Fellow,~IEEE,}
        and~Henry Arguello~\IEEEmembership{Senior Member,~IEEE}
\thanks{P. Goyes-Pe\~nafiel, C. V. Correa and H. Arguello are with the Department of Computer Science, Universidad Industrial de Santander, Bucaramanga, 680002 Colombia.}
\thanks{E. Vargas is with the Department of Electronics Engineering, Universidad Industrial de Santander, Bucaramanga 680002 Colombia.}
\thanks{Y. Sun is with the Department of Computer Science \& Engineering, Washington University in St.~Louis, St.~Louis, MO 63130, USA.}
\thanks{U.~S.~Kamilov is with the Department of Electrical \& System Engineering and the Department of Computer Science \& Engineering, Washington University in St.~Louis, St.~Louis, MO 63130, USA.}
\thanks{B. Wohlberg is with Theoretical Division, Los Alamos National Laboratory, Los Alamos, NM 87545 USA.}
}

\maketitle

\begin{abstract}
Physical and budget constraints often result in irregular sampling, which complicates accurate subsurface imaging. Pre-processing approaches, such as missing trace or shot interpolation, are typically employed to enhance seismic data in such cases. Recently, deep learning has been used to address the trace interpolation problem at the expense of large amounts of
training data to adequately represent typical seismic events. Nonetheless, most research in this area has focused on trace reconstruction, with little attention having been devoted to shot interpolation. Furthermore, existing methods assume regularly spaced receivers/sources failing in approximating seismic data from real (irregular) surveys. This work presents a novel shot gather interpolation approach which uses a continuous coordinate-based representation of the acquired seismic wavefield parameterized by a neural network. The proposed unsupervised approach, which we call coordinate-based seismic interpolation (CoBSI), enables the prediction of specific seismic characteristics in irregular land surveys without using external data during neural network training. Experimental results on real and synthetic 3D data validate the ability of the proposed method to estimate continuous smooth seismic events in the time-space and frequency-wavenumber domains, improving sparsity or low-rank-based interpolation methods.

\end{abstract}
\let\thefootnote\relax\footnotetext{$^*$ denotes equal contributions.}

%
\IEEEpeerreviewmaketitle

\section{Introduction}
%
%
%
%
\IEEEPARstart{I}{nterpolation} is of great importance within the seismic data processing workflow because environmental or topographic restrictions usually result in incomplete and irregular receiver and source sampling. Some limitations include natural and anthropological factors such as water bodies, and infrastructure, as well as equipment errors \cite{Cordsen2000}. The most common seismic interpolation approach involves recovering missing traces of a shot gather. A more complex approach focuses on estimating complete missing shot gathers, entailing greater economic, environmental, and implementation benefits.
Nonetheless, most of the work reported in the literature deals with trace interpolation, with limited consideration having been given to shot interpolation. Thus, this work focuses on interpolating missing seismic shots in irregularly sampled land surveys.

Shot interpolation has typically been addressed by convex optimization, aiming at inverting the acquisition model using a regularizer that imposes prior knowledge about the data such as sparsity in domains like wavelet, curvelet, shearlet and learned dictionaries \cite{Villarreal2018, Villarreal2019, Galvis2020}. 
More recently\cite{Goyes-Penafiel2021}, sparse regularization has been jointly integrated with implicit regularization provided by denoising algorithms by using the plug and play priors and consensus equilibrium framework \cite{Venkatakrishnan2013, kamilov2022pnp, Buzzard2018}.
The versatility of deep learning has also been explored to solve the shot interpolation problem in a supervised fashion \cite{Resnet_Interp}, specifically with a residual network architecture trained on data samples generated with bicubic interpolation of the incomplete data set.

A key aspect of supervised deep learning approaches for seismic data interpolation is that they employ external data sets. For instance, a number of authors \cite{Wang2020, Liu2021, Wang2019a, Fang2021, Zhang2020} employ convolutional neural networks (CNN) following an end-to-end training strategy that requires large training data sets. Alternatively, deep internal learning approaches that exploit the structural redundancy of the field data itself, rather than employing vast training data sets, have been proposed for seismic trace interpolation  \cite{Zhang_2019_ICCV, Shocher_2018_CVPR}, using deep image priors (DIP) \cite{park2020seismic, DIP_IrregTraces} and recurrent neural networks (RNN) \cite{RNN_TraceInterp}. Further, a combination of internal and external learning for trace interpolation has been studied\cite{Trace_InternalExternal}.
Although all these works have explored deep learning-based solutions for irregular sub-sampling schemes \cite{Mandelli2019, IrregularInterp_DL}, they implicitly require a binning process to rearrange irregularly sampled seismic data onto a regular grid with missing entries (traces). 

In contrast, this work presents a deep internal learning approach to estimate complete missing shot gathers in an irregular land survey bypassing the binning step. The proposed method takes advantage of a recent branch of work in computer graphics, coordinate-based neural representations, which allows the encoding of a continuous spatial field into the weights of a multilayer perceptron (MLP) by mapping coordinates to pixel values, in an unsupervised manner~\cite{Sun2021, Sitzmann2020, Mildenhall2020}. Specifically, the proposed coordinate-based seismic interpolation (CoBSI) method learns a continuous mapping from the spatial and temporal coordinates of the (incomplete) acquired seismic data to the underlying recorded field. The continuous nature of the neural representation can model irregular sampling scenarios without a binning process and is not constrained to have a spatial resolution, reducing memory costs compared with discrete representations. Furthermore, in contrast to current state-of-the-art methods, the proposed approach enables seismic shot interpolation for both regular and irregular 3D land surveys, in an unsupervised fashion, i.e., without additional training data. The proposed approach is validated using 3D seismic data from orthogonal grids such as cross-spreads, focusing specially on randomly subsampled regular and irregular grids with acoustic synthetic data, \textit{Stratton survey}~\cite{Stratton_3D_survey} and \textit{SEAM Phase II Foothills model}~\cite{Regone2017}. The results show that CoBSI outperforms the multichannel singular spectrum analysis (MSSA), damped-MSAA (DMSSA), sequential generalized K-means (SGK), and sparsity-based shot reconstruction methods.

\section{Background}

Coordinate-based neural representations have been successfully applied to unsupervised generation of highly realistic views of scenes with complicated geometries and appearance \cite{Mildenhall2020,Tancik2020}, and implicit neural representations of signals for solving boundary value problems \cite{Sitzmann2020}. In the same line of work, the coordinate-based internal learning (CoIL) approach in \cite{Sun2021} extrapolated these ideas to solve imaging inverse problems by modeling a continuous measurement field from a subsampled and noisy set of measurements, using geometry parameters of a tomographic imaging system. 
Since a seismic acquisition can be described in terms of a coordinate system, resembling the continuous field modeling from the computer vision area, this work explores a coordinate-based modeling to address the seismic shot interpolation problem in a deep internal learning approach. It is worth pointing out that CoBSI addresses a substantially more complex interpolation than that in \cite{Sun2021}, since the seismic wavefield we want to interpolate contains different responses from reflected, refracted, and surface waves.

This approach consists of two main processing blocks, a positional encoder and a multilayer perceptron (MLP). The positional encoding maps help to preserve high-frequency information through the encoding of coordinate positions \cite{02561}, while the MLP works as an interpolator from the encoded coordinates to the signal amplitude. For instance,  \cite{Tancik2020} shows that passing input points through a simple Fourier feature mapping enables a MLP to learn high-frequency functions in low-dimensional problem domains. It is worth noting that the seismic interpolation task with MLP relies on a low-dimensional problem as explained below.  

\section{3D Seismic Acquisition Model}

\begin{figure}[ht]
    \centering
    \includegraphics[width=\columnwidth]{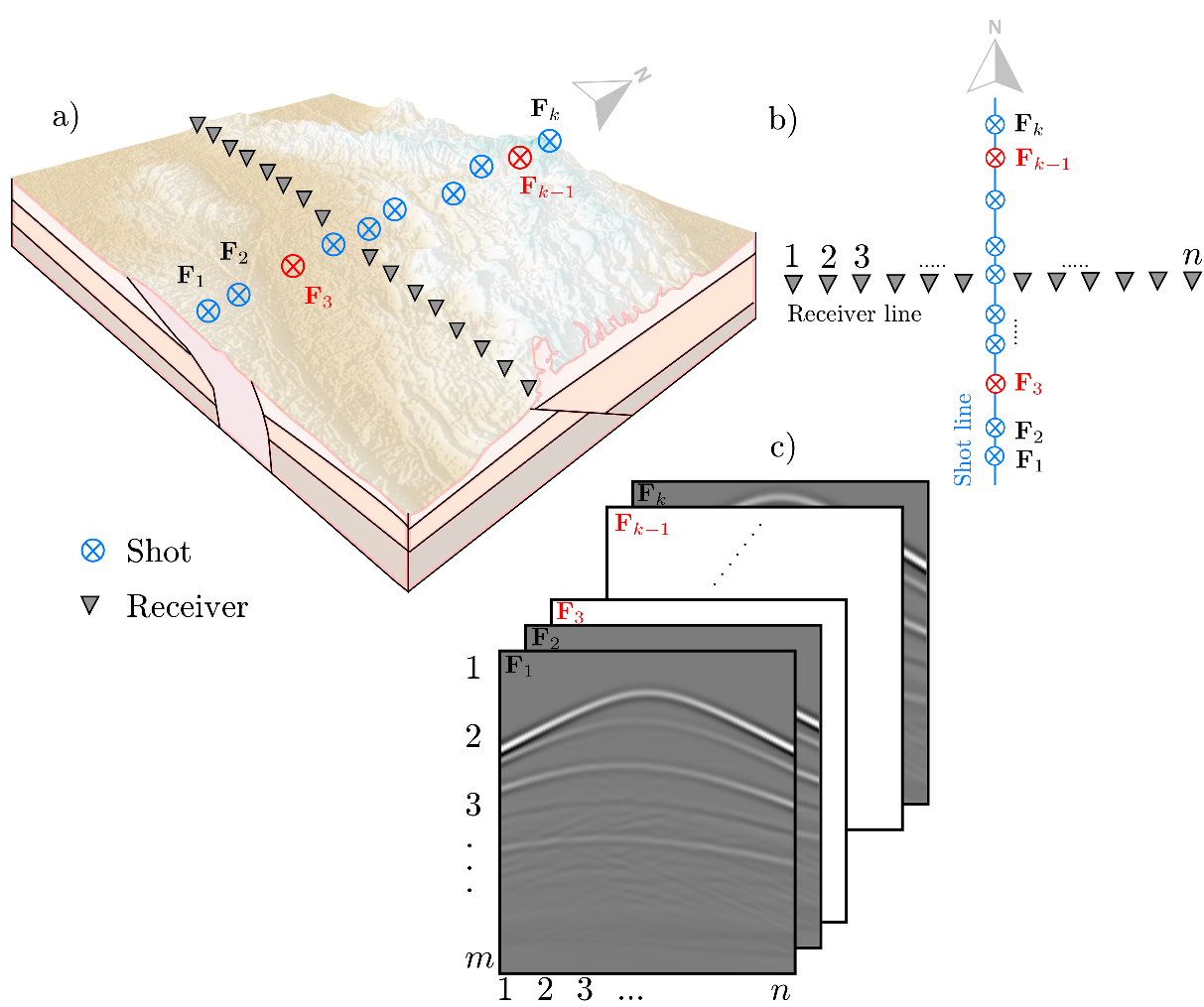}
    \caption{Cross-spread geometry in irregular survey with missing shots $\mathbf{F}_3$ and $\mathbf{F}_{k-1}$ in red. (a) Perspective view. (b) Plan view and (c) Seismic data in tensor representation $\mathcal{F}$.}
    \label{fig:Cross_spread_Scheme2}
\end{figure}

Ideal seismic surveys are orthogonal grids with uniformly spaced receivers and sources (i.e. pre--plot design). In practice however, environmental and topographic restrictions induce irregularities that result in non-uniform spatial intervals, as illustrated by the cross-spread acquisition example in Figure~\ref{fig:Cross_spread_Scheme2}(a-b), where missing shots are depicted in red. Data from cross-spread surveys is modeled as cubes of $k$ stacked shot gathers $\mathbf{F}_i \in \mathbb{R}^{m \times n}, $ with $m$ time samples and $n$ receivers. Thus, the whole data set can be denoted as a tensor $\mathcal{F}= \{ \mathbf{F}_i\}_{i=1}^{k} \in \mathbb{R}^{m\times n \times k}$. Figure ~\ref{fig:Cross_spread_Scheme2}(c) shows a survey with missing shot gathers $\mathbf{F}_3$ and $\mathbf{F}_{k-1}$. Letting $\mathbf{f} \in \mathbb{R}^{mnk}$ be the vector representation of the full cross-spread seismic survey $\mathcal{F}$, the acquisition model can be written as the linear system 
\begin{equation}
    \mathbf{r}=\mathbf{\Phi}\mathbf{f}+\omega  \,,
    \label{eq:linearmodel}
\end{equation}
with $\mathbf{\Phi}$ as the matrix modeling the sampling process, $\omega$ the acquisition noise, and $\mathbf{r} \in \mathbb{R}^{mn(k-s)}$ is the incomplete acquired data (seismic response). Specifically, the acquisition operator $\mathbf{\Phi} \in \mathbb{R}^{mn(k-s)\times mnk}$ is defined as $\mathbf{\Phi}=\mathbf{S}\otimes \mathbf{I}_{mn}$, where $\otimes$ represents the Kronecker product\cite{kronProd}, $\mathbf{I}_{mn}$ is a $mn\times mn$ identity matrix, $\mathbf{S}\in \mathds{R}^{k-s\times k}$ is an identity matrix modeling the subsampling effect by setting to zero the $s$ rows that correspond to the linear indices of the missing sources.

\begin{figure}[ht]
    \centering
    \includegraphics[width=0.8\columnwidth]{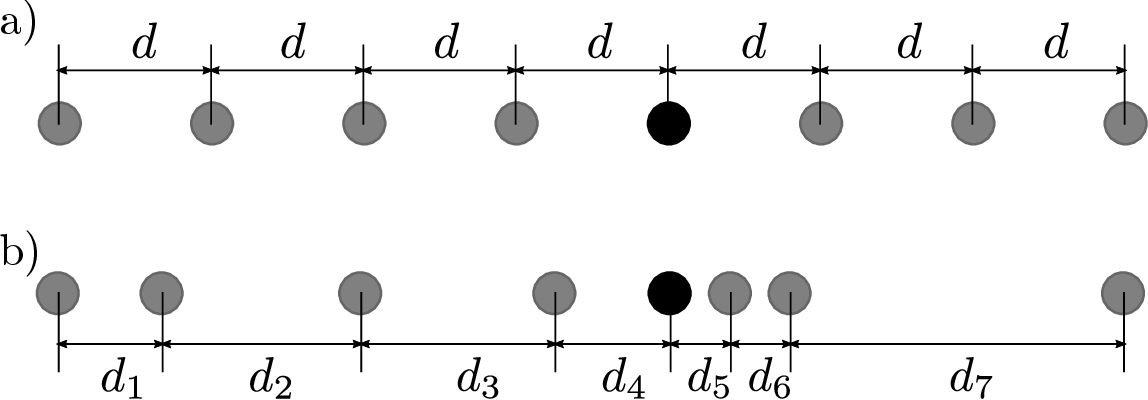}
    \caption{Source sampling with (a) uniform and regular interval with distance $d$ and, (b) irregular interval with variable distance. Black dots represent missing shots to be interpolated.}
    \label{fig:irregularShooting}
\end{figure}

Previous works have shown that the underlying seismic data $\mathbf{f}$ can be estimated from the incomplete acquisition $\mathbf{y}$, following either optimization or data-driven approaches. Specifically, optimization methods consider that seismic signals are sparse in some transformation domain $\mathbf{\Psi}$, such that they can be represented as $\mathbf{\Psi} \mathbf{f} =  \alpha$, where $\alpha$ corresponds to the coefficients in the transformed domain \cite{Goyes-Penafiel2021, Villarreal2018, Villarreal2019, Herrmann2012a, Li2012}. Using such a sparsity prior, it is possible to estimate $\mathbf{f}$ by solving the optimization problem given by 

\begin{equation}
    \mathbf{f}^* = \argmin_{\mathbf{f}} \; (1/2) \left \| \mathbf{r} - \mathbf{\Phi} \mathbf{f} \right \|_{2}^{2} + \lambda \left \| \mathbf{\Psi}\mathbf{f} \right \|_{1} \,,
\label{eq:sparseprior}
\end{equation}
where $\lambda>0$ is a regularization parameter weighting the sparsity term in the solution. On the other hand, missing seismic data can be recovered using data-driven approaches, such as convolutional neural networks that learn internal structures of extensive seismic data sets \cite{Liu2021, Wang2020, Wang2019a}. Nonetheless, all these methods rely on the sensing matrix $\mathbf{\Phi}$, which accounts for indexed sampling positions. Thus, it assumes a regular sampling grid, as illustrated in Figure \ref{fig:irregularShooting}(a), i.e., it cannot model irregularly-spaced sources. Therefore, reconstruction methods still face limitations in providing accurate seismic estimates for irregular surveys, as the one depicted in Figure \ref{fig:irregularShooting}(b), where the distance between sources is not fixed.

To address this problem,~\cite{Hennenfent2010} incorporated a non-equispaced curvelet transform within a sparsity-promoting prior, and~\cite{Galvis2020} used interpolation operators on the irregular grid before applying reconstruction algorithms with a binning pre-processing step to cast the irregularly sampled data into a regular grid. The main drawback of these approaches is that the interpolator assumes linear continuity, since it is applied only in the shots direction, which in 3D acquisitions does not allow interpolation of the two-dimensional wavefield. For this reason, the approach in~\cite{Galvis2020} is limited to 2D shots in split-spread geometries.

\section{Coordinate-based Seismic Shot Interpolation}

\begin{figure}[ht]
    \centering
    \includegraphics[width=0.7\columnwidth]{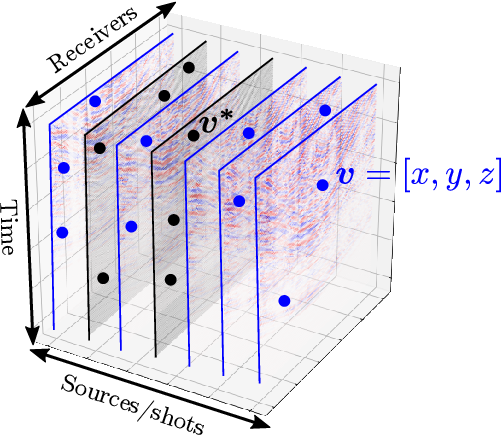}
    \caption{Illustration of the coordinate-based representations $\boldsymbol{v}$ in a single cross-spread grid for seismic acquisition (blue dots), with $x,y,z$ denoting receiver, time, and source coordinates, respectively. Coordinates $\boldsymbol{v}^{*}$ belong to the desired shot gathers to interpolate (black dots).}
    \label{fig:cross-spread}
\end{figure}

Unlike state of the art interpolation methods that rely on index-based modeling of the survey, the proposed coordinate-based seismic interpolation (CoBSI) method employs a coordinate-based deep internal learning approach for modeling the seismic survey in a more realistic fashion. The core idea of the proposed approach is motivated by recent computational imaging works in neural interpolation \cite{Sun2021, Barron2021, Sitzmann2020, Mildenhall2020, Wang2020a, Tancik2020}, and consists in representing the acquired response of the wavefield $r \in \mathbb{R}$ from a given coordinate $\boldsymbol{v} = [x,y,z] \in \mathbb{R}^3$, where $x,y,z$ respectively denote the time, receiver and source positions, with a neural network $\mathcal{M}_{\boldsymbol{\theta}}$ with parameters ${\boldsymbol{\theta}}$. The goal of this neural network is to map input coordinates to the sampled wavefield responses, i.e.,  $r = \mathcal{M}_{\boldsymbol{\theta}}(\boldsymbol{v})$. Based on this representation, we can model the acquired  cross-spread $\mathbf{r}$ from Equation \eqref{eq:linearmodel} by querying $\mathcal{M}_{\boldsymbol{\theta}}$ using the \newtext{corresponding} coordinates \newtext{of the acquired response} (see Figure \ref{fig:cross-spread}).

The proposed neural network $\mathcal{M}_{\boldsymbol{\theta}}$ is the composition of a high dimensional mapping $\gamma$, and a multilayer perceptron (MLP) $\mathcal{N}_{\boldsymbol{\theta}}$ such that $\mathcal{M}_{\boldsymbol{\theta}}(\boldsymbol{v}) = \mathcal{N}_{\boldsymbol{\theta}}(\gamma (\boldsymbol{v}))$. Recent works have demonstrated that this separation mitigates the performance degradation observed in MLP to represent high frequency variations~\cite{Tancik2020, Barron2021, Sun2019}, such as those measured in seismic data due to abrupt changes in the continuity of reflection events, or coherent noise such as head waves and ground roll. The following subsections  present the details of the mapping function $\gamma$ and the MLP  $\mathcal{N}_{\boldsymbol{\theta}}$ architecture. 

\subsection{Anisotropic Positional Encoding}
\label{sec:pe}
To address the problem of representing high frequency components of natural images,~\cite{Tancik2020} proposed to employ a positional encoding $\gamma$ as high dimensional mapping, given by

 \begin{multline}
    \gamma_{U}(\boldsymbol{v})=[ \cos(\omega_1 \boldsymbol{v}), \sin(\omega_1 \boldsymbol{v}),...,
    \cos(\omega_i \boldsymbol{v}), \sin(\omega_i \boldsymbol{v}),..., \\
    \cos(\omega_U \boldsymbol{v}), \sin(\omega_U \boldsymbol{v})]^\mathrm{T} \,, \hspace{0.1cm}
    \label{eq:positionalencoding}
\end{multline}
where $U$ is the total number of components, $\{ \omega_i \}_{i=1}^{U}$, the frequency mappings are given by $\omega_i = i\pi/2$ or $\omega_i = \pi 2^{i-1}$ in the linear and exponential sampling, respectively, and $\boldsymbol{v}$ is some arbitrary coordinate normalized to lie in $[0,1]^3$. Note that the positional encoding defined in Equation \eqref{eq:positionalencoding} expands the input coordinates as the combination of different frequency components and, all coordinates are mapped to the same number of frequencies.

\begin{figure}[ht]
    \centering
    \includegraphics[width=0.8\columnwidth]{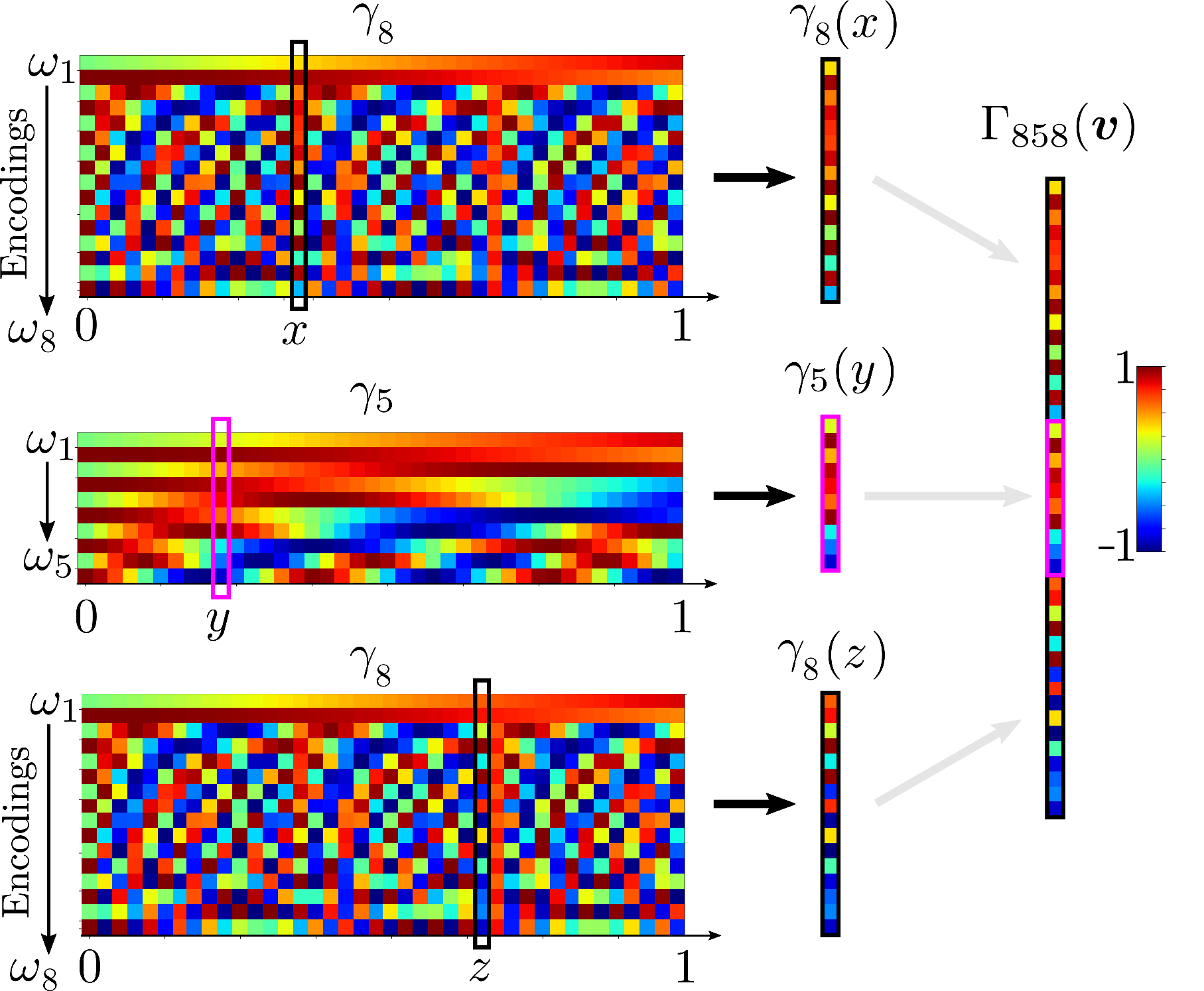}
    \caption{Conceptual example of anisotropic positional encoding maps with linear sampling with $M=8$, $N=5$ and $K=8$ number of frequencies. The matrices on the left represent the $\gamma$ functions for each coordinate, respectively. The vectors in the middle represent the corresponding mapping for a given coordinate $x$, $y$ and $z$. Finally, the vector in the right $\Gamma_{858}(\boldsymbol{v})$ illustrates the final mapping of the proposed anistropic positional encoding.}
    \label{fig:encoding3}
\end{figure}

However, in the particular case of seismic data, each signal coordinate represents substantially different features (sources, receivers and time samples), which should not be equally encoded to preserve the structure of the data. Therefore, this work proposes to employ an anisotropic positional encoding, in which a different number of frequency components is used for each axis direction. Thus, for the particular three dimensional case of seismic data, the anisotropic positional encoding is defined as

\begin{equation}
\Gamma_{MNK}(\boldsymbol{v})=[ \gamma_{M}(x),\gamma_{N}(y),\gamma_{K}(z) ]^T \hspace{0.1cm},
\label{eq:gammencoding}
\end{equation}
where $M,N,K$ are the number of frequency components associated with 
$x$ (time), $y$ (receiver) and $z$ (shot) axes as shown in Figure  \ref{fig:cross-spread}.
Figure~\ref{fig:encoding3} illustrates an example of the anisotropic positional encoding for $M=8$, $N=5$ and $K=8$, where the horizontal axis for all encoding maps represents the normalized coordinate values lying in $[0,1]$, the vertical axis represents the number of encoding frequencies, and the output $\Gamma$ lies in $[-1,1]$. The number of frequency components for each coordinate $M,N,K$ is found via parameter tuning.

\subsection{Network Architecture}
\label{sec:network}
A multilayer perceptron (MLP) is here used to approximate the function $\mathcal{N}_\theta$. This architecture is modeled as the nested functions

\begin{equation}
    \label{eq:MLP}
    \mathcal{N}_{\boldsymbol{\theta}} (\boldsymbol{v}) = f_L\left(f_{L-1}\left(\cdots f_2(f_1(\boldsymbol{v}))\right)\right) \,,
\end{equation}
where $L$ is the number of layers or depth of the MLP, and
$f_i( \boldsymbol{v}_i)= \phi\left(\mathbf{W}_i\boldsymbol{v}_i + \mathbf{b}_i\right)$
is the $i^{\mathrm{th}}$ layer of the MLP,  which is an affine transformation represented by the matrix $\mathbf{W}_i$ and bias $\mathbf{b}_i$, followed by a non-linear activation function $\phi$. This work adopts the rectified linear unit (ReLU) as activation function, which is one of the most widely employed in modern neural networks \cite{nair2010rectified}. Moreover, since seismic data are normalized in the range $[0,1]$, the sigmoid activation function has been selected for the output layer.

\begin{figure}[ht]
    \centering
    \includegraphics[width=0.8\columnwidth]{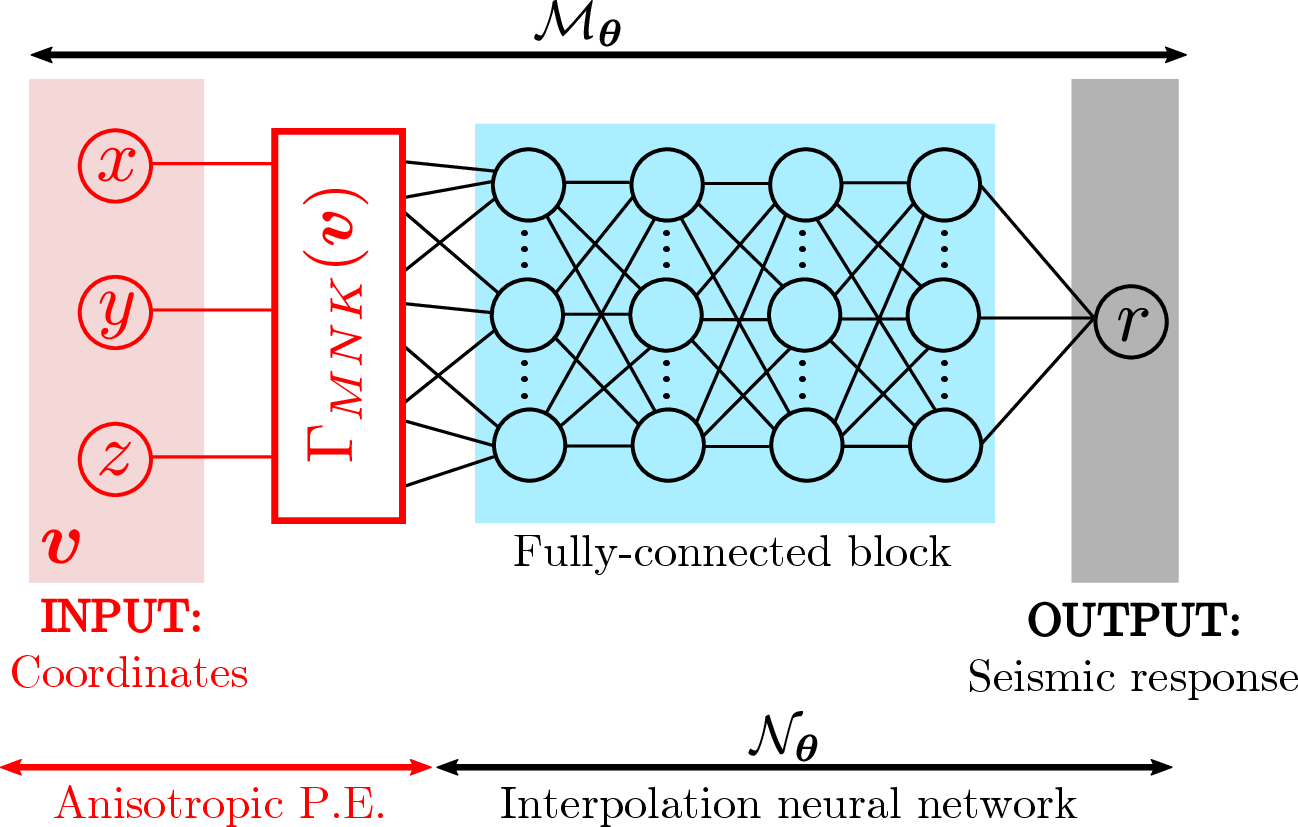}
    \caption{Coordinate-based neural network $\mathcal{M}_{\boldsymbol{\theta}}$ with anisotropic Positional Encoding (P.E) and interpolation neural network $\mathcal{N}_{\boldsymbol{\theta}}$}
    \label{fig:ANN}
\end{figure}

\newtext{The overview of the proposed  coordinate-based neural network is depicted in Figure~\ref{fig:ANN} with a fully-connected block of $\mathrm{NN}$ neurons and $L$ layers, and a single-neuron output layer}. Note that it can be seen as a regression problem, such that we can simply employ a mean squared error (MSE) loss function to find the optimal network parameters, $\boldsymbol{\theta^*}$.  The MSE loss function can be written as
\begin{equation}
\mathcal{L}(\boldsymbol{\theta}) = \frac{1}{mn(k-s)} \sum_{i}^{mn(k-s)} ( r_i - \mathcal{N}_{\boldsymbol{\theta}} (\Gamma_{MNK}(\boldsymbol{{v}}_i)) )^2 \,,
    \label{eq:mlploss}
\end{equation}
\newtext{where $s$ and $k$ are the number of missing and total shots, respectively; $\boldsymbol{v}_i=[x_i,y_i,z_i]$ is the $i^{\mathrm{th}}$ element of the set of coordinates $\mathcal{V}$, 
and $r_i$ is the $i$-th entry of the acquired amplitude values $\mathbf{r}$ from Equation \eqref{eq:linearmodel}.}

We remark here that training data from additional seismic surveys are not necessary because we just employ the available discrete samples of the acquisition of interest for training the network.

\newtext{\subsection{Interpolation Algorithm}}

\newtext{
The anisotropic positional encoding and MLP from sections \ref{sec:pe} and \ref{sec:network} are combined as the CoBSI method decribed in Algorithm \ref{alg:cobsi}. 
Specifically, the inputs of the algorithm are:
}

\begin{itemize}

\item[i)] \newtext{The set of coordinates of the available (acquired) data $\mathcal{V}= \left \{ \boldsymbol{v}_i \right \}_{i=1}^{mn(k-s)}$, with $\boldsymbol{v}_i=[x_i,y_i,z_i]$}

\item[ii)] \newtext{The corresponding amplitude values  of the acquired data $\mathbf{r} =[r_1, \cdots , r_i, \cdots, r_{mn(k-s)}]$ from Eq. \eqref{eq:linearmodel}. }
\item[iii)] \newtext{The set of coordinates of the missing data $\mathcal{V}^*= \left \{ \boldsymbol{v}_{i}^{*} \right \}_{i=1}^{mns}$}

\end{itemize}

\newtext{In step 1, the anisotropic positional encoding of each point $\boldsymbol{v}_i$ is calculated using Eq. 4. Then, in lines $3-10$ the positional encodings and the amplitude values are used to train the MLP $\mathcal{N}_{\theta}$ in an end-to-end fashion, to obtain the optimized network parameters $\pmb{\theta^*}$, as in step $11$ of Algorithm \ref{alg:cobsi}. In lines $12-14$, the network with optimal parameters is used to estimate the amplitude values of the missing shots by querying $\mathcal{M}_{\boldsymbol{\theta}}$, using the corresponding coordinates $\mathcal{V}^* =  \lbrace \boldsymbol{v}_i^{*} \rbrace_{i=1}^{mns} $ of the  $s$ missing shot-gathers, 
as long as the input coordinates are in the same acquisition domain. Specifically, note that in terms of seismic surveys, the acquisition domain is related to the maximum coverage in receiver and source lines as shown in Figure~\ref{fig:cross-spread}. In Line $15$, these amplitudes are concatenated to the known data amplitudes, and rearranged as 2D structures corresponding to the seismic shots (Line $16$). Thus, a full seismic data cube $\mathcal{F}$ is obtained.}



\begin{algorithm}[ht]
\begin{algocolor}
    \caption{\newtext{Seismic shot interpolation using CoBSI method}}\label{alg:cobsi}
    \begin{algorithmic}[1]
        \Require $\mathcal{V}$: Set of coordinates; $\mathbf{r}$: Data amplitude values corresponding to each coordinate. $\mathcal{V}^*$: Set of coordinates for the missing shots. $\mathrm{N}_\mathrm{E}$: Number of iterations. 

        \State   
        Calculate set $\mathcal{P} =  \left \{ \Gamma_{MNK}(\boldsymbol{v}_i) \right \}_{i=1}^{mn(k-s)}$ using Eq. 4 and $\boldsymbol{v}_i \in \mathcal{V}$ . %
        
        \State Initialize $\boldsymbol{\theta}$ randomly
        \For {\text{ $i=1....\mathrm{N}_\mathrm{E}$}}
            \State Draw $\mathcal{P}_t \subset \mathcal{P}$, $\mathbf{r}_t$ from $\mathbf{r}$ \Comment{Draw data batch}
            \For{\text{each $\Gamma_{MNK}(\boldsymbol{v}_j) \in \mathcal{P}_t$, $\mathbf{r}_j $ from $\mathbf{r}_t$  }}
            \State $\hat{{r}}_j \gets \mathcal{N}_{\theta}(\Gamma_{MNK}(\boldsymbol{v}_j) )$ \Comment{Estimate the seismic response}
            \State Compute MSE loss $\mathcal{L}(\boldsymbol{\theta})$ using $\hat{{r}}_j,r_j$, and Eq. 6. 
            \State Update $\boldsymbol{\theta}$ using ADAM optimizer
            \EndFor
        \EndFor
        \State Get the optimal parameter $\boldsymbol{\theta}^*$ from last iteration
        \For{\text{each $\boldsymbol{v}_i^* \in \mathcal{V}^*$}}        
            \State $\hat{{r}}_i^* \gets \mathcal{N}_{\boldsymbol{\theta}^*}(\Gamma_{MNK}(\boldsymbol{v}_i^*) )$ \Comment{Estimate the seismic responses for missing shots}
        \EndFor
        \State $\mathbf{f} \gets [  \hat{\mathbf{r}}^*, \mathbf{r} ] $ \Comment{Concatenate acquired and interpolated seismic response}
        \State $\mathcal{F} \in \mathbb{R}^{m\times n \times k} \gets \text{reshape}(\mathbf{f} \in \mathbb{R}^{mnk})$ \Comment{Rearrange vector to tensor representation}
        \State \textbf{Output}: $\mathcal{F}$ complete seismic data
    \end{algorithmic}
    \end{algocolor}
    \end{algorithm}

\section{Simulations and Results}

Three different experiments were carried out to evaluate the effectiveness of CoBSI in regular and irregular cross-spread acquisition geometries, using synthetic and real data. In all the experiments, we fixed the number of layers of the MLP architecture to $L=15$ with the same number of neurons (NN) per layer. Table~\ref{tbl:summ} summarizes the main network parameters used for training on each experiment: data set, NN, learning rate, epochs, and number of trainable parameters (NTP). \newtext{These parameters were found by grid search, so that the best PSNR metric was obtained on each experiment.}

\begin{table}[ht]
    \caption{Summary of parameters for the coordinate-based neural network $\mathcal{M}_{\boldsymbol{\theta}}$ on each experiment. NN: Number of neurons, LR: Learning Rate, NTP: Number of trainable parameters.}
    \label{tbl:summ}
    \centering
\begin{tabular}{cc|cccc}
\hline
Experiment & Dataset &  NN & LR & Epochs & NTP \\ \hline
\textbf{I}  & Synthetic &         $128 $            & 1e-3 & 1000  & 232449 \\ 
\textbf{II} & Stratton       & $256$                & 1e-3 & 5000  & 932865   \\ 
\textbf{III}  & SEAM Phase II       & $128$          & 1e-4 & 50000 &  234241  \\ \hline
\end{tabular}
\end{table}

Particular details on each data set and experiment are included in the following subsections. All experiments were conducted using a NVIDIA Tesla P100 16GB GPU. The peak signal-to-noise ratio (PSNR) is here used to assess the accuracy of the reconstructions, exactly as described by \cite{Liang2014}, as well as the Structural Similarity Image Metric (SSIM) from \cite{Wang2004}. Both metrics were applied in the time-domain shots, with respect to the ground truth. Interpolated shots obtained with CoBSI are compared against those resulting from the F-XY domain multichannel singular spectrum analysis (MSSA) and damped-MSSA (DMSSA) method \cite{huang2016dmssa} with fhigh=550 and iter=50 implemented using the DRR Matlab package~\cite{chen2016drr3d}, sequential generalized K-means (SGK) using fast dictionary learning for high-dimensional seismic reconstruction~\cite{chen2020sgk, wang2021sgk}, and the sparsity-based interpolation (SBI) in Equation \eqref{eq:sparseprior}, solved with the ADMM algorithm from~\cite{Villarreal2019}. \newtext{All the parameters in these methods were fixed to those suggested by the authors of each work.}



\begin{table*}[ht]
    \caption{Results summary for the interpolated shots obtained with CoBSI, compared with respect to DMSSA, MSSA, SGK and SBI methods using three different seismic acquisition surveys. }
    \label{tbl:exps}
    \centering
\begin{tabular}{lc|clllc|clllc}
\hline
\multirow{2}{*}{Experiment}            & \multirow{2}{*}{Shot} & \multicolumn{5}{c|}{SSIM}             & \multicolumn{5}{c}{PSNR (dB)}         \\ 
                                       &                       & CoBSI & DMSSA & MSSA & SGK & SBI & CoBSI & DMSSA & MSSA & SGK &  SBI \\ \hline

 & S4 & 0.990 & 0.294 & 0.284 & 0.493 & 0.547 & \textbf{42.368} & 17.737 & 18.379 & 19.261  & 20.845    \\ \cline{2-2}
 \multirow{3}{*}{Experiment I: }  & S6                    & \textbf{0.983}  &  0.237      &  0.239    & 0.459   & 0.741     & \textbf{42.070} &   14.933   &   15.707    &   19.146   & 26.463   \\ \cline{2-2} 
\multirow{3}{*}{\textbf{Synthetic}}  & S8                    & \textbf{0.977}  &  0.277      &   0.253    &  0.453   & 0.921     & \textbf{36.589} &   20.276    &   19.486   &  19.562    & 22.026   \\ \cline{2-2}          
    & S11                    & \textbf{0.977}  &  0.183     & 0.135     &    0.461 & 0.583     & \textbf{37.343} & 19.241      &  17.867    &  18.155   & 21.199    \\ \cline{2-2}
                                        & S13                    & \textbf{0.980}  &  0.060     &   0.065    &    0.431 & 0.481     & \textbf{37.479} &   14.188    &  14.318    &  15.609   & 17.264    \\ \cline{2-12} 
                                       & Average               & \textbf{0.981}   &   0.210   &   0.195    &  0.459   & 0.655   & \textbf{39.170} &   17.275    &   17.151    &  18.347   & 21.559  \\ \hline
\multirow{3}{*}{Experiment II:} 
& S3                    & \textbf{0.651}  &   0.084     &  0.070     &  0.435   & 0.503     & \textbf{22.621} &  15.419     &   14.850   &  16.510   & 17.371    \\ \cline{2-2}
\multirow{3}{*}{\textbf{Stratton 3D survey}}  & S6                    & \textbf{0.552}  &  0.106     & 0.092     &  0.402   & 0.520     & \textbf{19.970} &  15.412      &  14.900    &   16.098  & 19.233    \\ \cline{2-12} 
 & Average               & \textbf{0.602}  &   0.095    &   0.081   &   0.419  & 0.512     & \textbf{21.296} &   15.416     &   14.875    &  16.304  & 18.302    \\ \hline
\multirow{3}{*}{Experiment III:} & S2                    & \textbf{0.724}  &   0.150    &   0.083    &  0.399   & 0.296     & \textbf{23.379} &    17.398    &   15.416    &  18.082   &  20.195    \\ \cline{2-2}
\multirow{3}{*}{\textbf{SEAM Phase II}}       & S4     & \textbf{0.728}  &   0.255     &   0.185    &  0.436    & 0.229     & \textbf{24.812} &   19.768    &   18.154   & 19.183    & 19.807   \\ \cline{2-12} 
                                       & Average               & \textbf{0.726}  &   0.203  &  0.134   &  0.418     & 0.263     & \textbf{24.096} &   18.583    &   16.785   &  18.633   & 20.001    \\ \hline
\end{tabular}

\end{table*}

\begin{figure}[ht]
    \centering
    \includegraphics[width=\columnwidth]{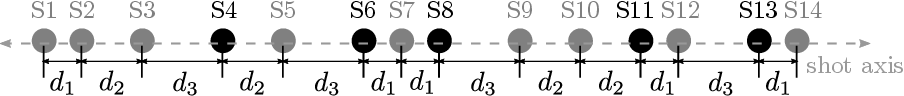}
    \caption{Synthetic irregular shot acquisition employed in Experiment I, with $d_1=75$ m, $d_2=100$ m and $d_3=125$ m. Black dots indicate missing shots.}
    \label{fig:scenario1}
\end{figure}

\textbf{Experiment I:} A synthetic data set of an irregular acquisition from a cross-spread grid using the acoustic forward modeling operator from DEVITO \cite{Louboutin2019,Luporini2018} to propagate the seismic wavefield was used in this experiment. The cross-spread comprises $m=900$ time samples, $n=101$ receivers and $k=14$ shots. The interval samplings are $dt=1$ ms, $dg = 25$ m, respectively for time and receivers. For this experiment, 5 shots (S4, S6, S8, S11, and S13) were removed from the data set, and interpolated using the proposed CoBSI method.  Figure~\ref{fig:scenario1} depicts the irregular interval in shot sampling, where the 5 missing shots are represented by black dots. Note that the removed shots account for different distance intervals among known data.

\begin{table}[ht]
    \caption{Interpolation results for $5$ missing shots in Experiment I, using the proposed CoBSI method with exponential and linear sampling, (CoBSI-Exp and CoBSI-Lin, respectively) }
    \label{tbl:exp1}
    \centering
\begin{tabular}{c|cc|cc}
\hline
\multirow{2}{*}{Shot} & \multicolumn{2}{c|}{SSIM}  & \multicolumn{2}{c}{PSNR (dB)} \\
                      & CoBSI-Exp      & CoBSI-Lin & CoBSI-Exp  & CoBSI-Lin        \\ \hline
S4                    & 0.990          & 0.987     & 42.368     & 42.701           \\
S6                    & 0.983          & 0.980     & 42.070     & 41.542           \\
S8                    & 0.977          & 0.969     & 36.859     & 34.767           \\
S11                   & 0.977          & 0.976     & 37.343     & 40.587           \\
S13                   & 0.980          & 0.977     & 37.479     & 36.707           \\ \hline
Average               & \textbf{0.982} & 0.978     & 39.224     & \textbf{39.261}  \\
Std. Dev.             & 0.005          & 0.007     & 2.746      & 3.373            \\ \hline
\end{tabular}

\end{table}

\begin{figure}[ht]
    \centering
    \includegraphics[width=\columnwidth]{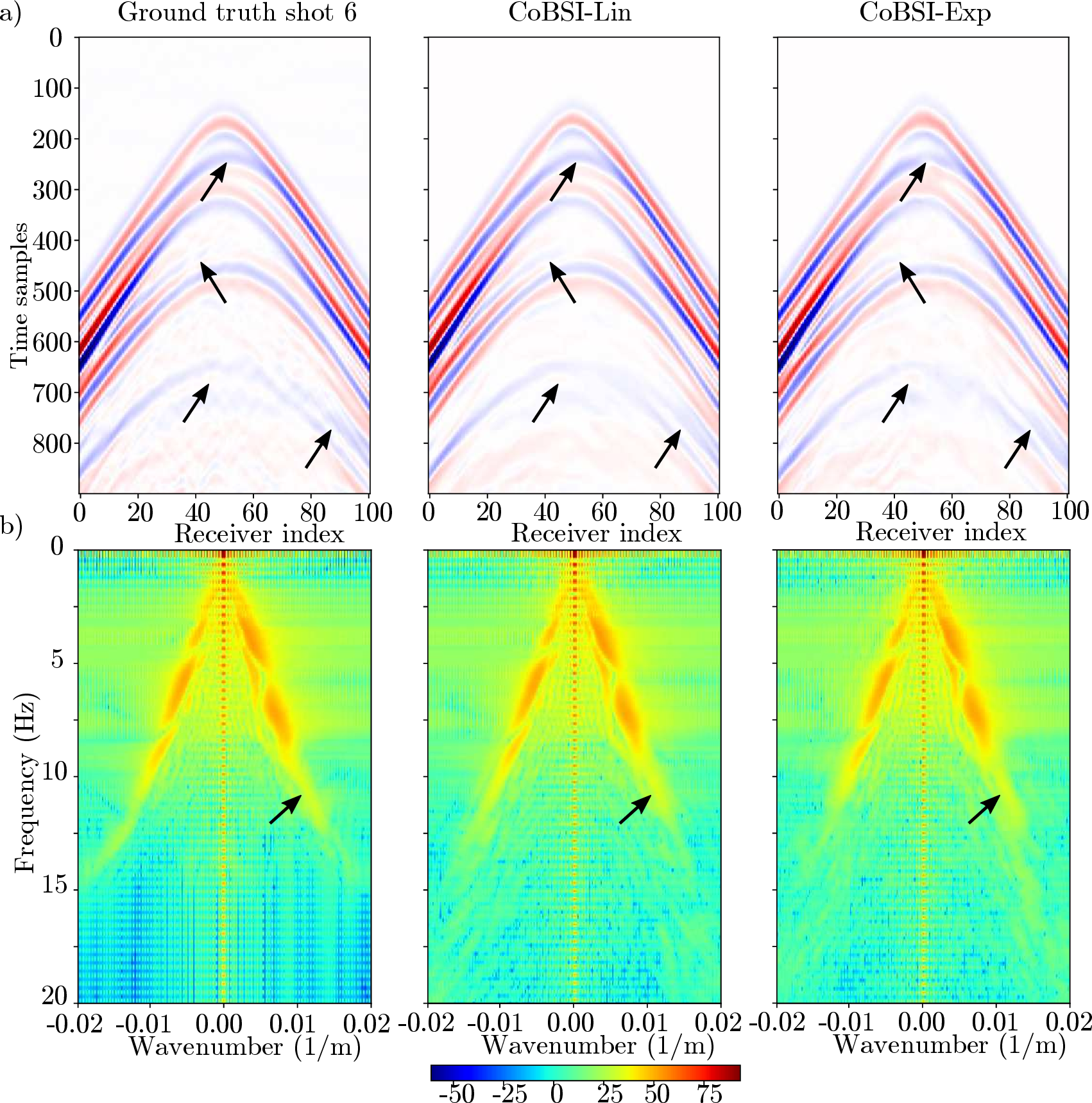}
    \caption{Comparison between Shot 6 interpolation using linear and exponential sampling in the positional encoding function for the experiment I in a) time and (b) frequency-wavenumber domain. Arrows point at the main differences and correspond to events whose continuity is better approximated by CoBSI-Lin.}
    \label{fig:E1shot_6mapping}
\end{figure}

Besides evaluating the ability of CoBSI to interpolate seismic shots from a synthetic irregular data set, this experiment  aims at analyzing the behavior of linear and exponential frequency mappings in the anisotropic positional encoding.  Table~\ref{tbl:exp1}  summarizes the CoBSI interpolation metrics for the 5 missing shots. Specifically, as a result of parameter tuning, the number of frequencies were fixed at 1 for time and shots, and 2 for receivers, i.e., $\Gamma_{121}$. The attained results show small variations in the metrics obtained with the two types of sampling for the anisotropical positional encoding function. Figure~\ref{fig:E1shot_6mapping} presents the interpolation results of shot S6, which exhibits discontinued seismic events highlighted by the arrows, that are not fully interpolated by CoBSI with exponential sampling. On the other hand, with linear sampling, CoBSI is able to smooth the entire signal by estimating more continuous events throughout the shot.

\begin{figure*}[ht]
    \centering
    \includegraphics[width=1\textwidth]{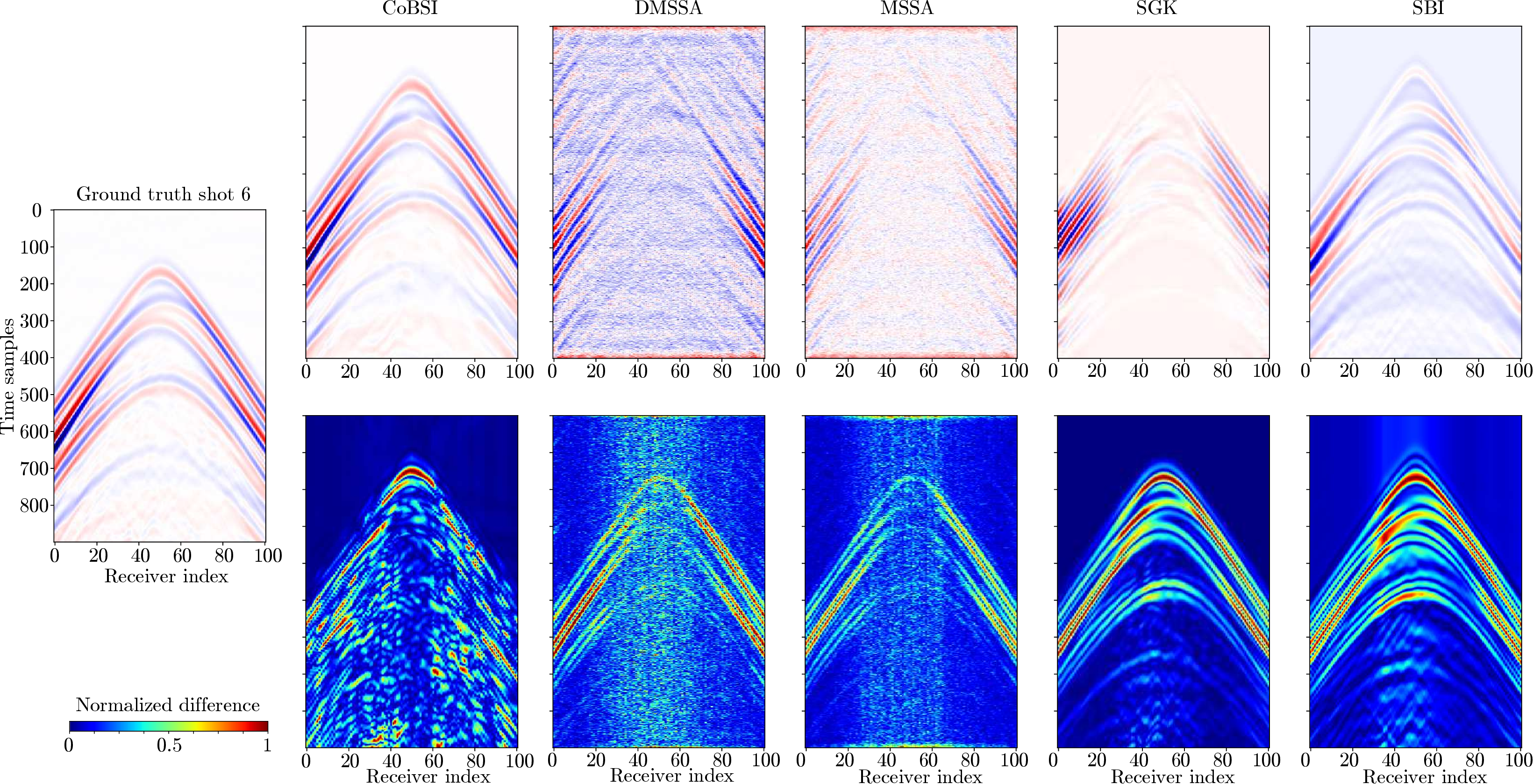}
    \caption{Comparison of Shot 6 interpolation results from Experiment I using CoBSI, DMSSA, MSSA, SGK and SBI methods.}
    \label{fig:E1shot_6}
\end{figure*}

Table~\ref{tbl:exps} summarizes CoBSI-Exp interpolation results compared with those from DMSSA, MSSA SGK and SBI. It can be seen that the proposed approach outperforms the evaluated counterparts in both metrics. Figure~\ref{fig:E1shot_6} shows that unlike the comparison methods, where most of the reconstruction errors occur in the reflection events, CoBSI shows a small distribution of errors towards the edges of the shot gather, thus, exhibiting a high reconstruction accuracy in the first arrivals, as well as in linear and hyperbolic events.  Moreover, considering the averaged results, CoBSI outperforms the comparative methods in up to 0.786 (SSIM) and 22.018 dB (PSNR).

\begin{figure}[ht]
    \centering
    \includegraphics[width=0.9\columnwidth]{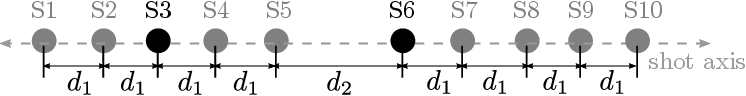}
    \caption{Irregular real seismic acquisition employed in Experiment II, with $d_1=50$ m, $d_2=100$  m. Black dots indicate missing shots.}
    \label{fig:scenario2}
\end{figure}

\textbf{Experiment II:} The seismic data set employed for this experiment is the \textit{Stratton survey}~\cite{Stratton_3D_survey}, a real 3D land swath acquisition project from South Texas, which was rearranged as a cross-spread using a geometric analysis based on the survey characteristics. A subset of 1001 time samples, 90 receivers, and 10 sources, with a gap between the fifth and sixth shots, as illustrated in Figure~\ref{fig:scenario2}. 
The goal of this experiment is to evaluate the ability of the proposed method to deal with real more complex data. To this end, we removed shots S3 and S6, such that two different gap lengths are considered. The interpolation challenge in these real data is to determine the  correct position of the seismic reflection event in the vertical axis (time axis). In this case, the number of frequencies for the positional encoding were fixed at 9 for the time axis, 5 for receivers and 8 for shots, i.e., $\Gamma_{958}$. 

\begin{figure*}[ht]
    \centering
    \includegraphics[width=1\textwidth]{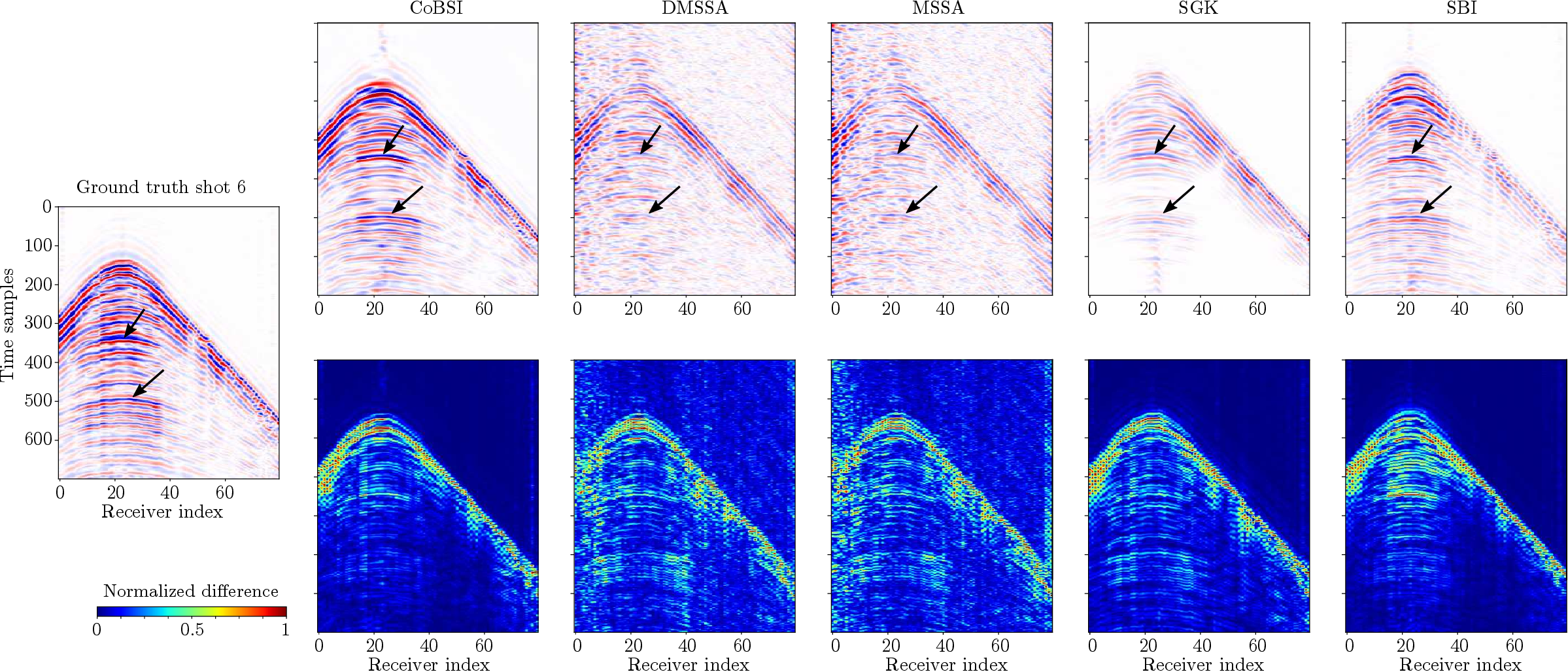}
    \caption{Interpolation results for Shot S6 from Experiment II, using CoBSI, DMSSA, MSSA, SGK and SBI methods. The arrows point to the main seismic reflection events. }
    \label{fig:E2R}
\end{figure*}

Figure~\ref{fig:E2R} presents a comparison between the interpolation results for shot S6, where it can be seen that the interpolated signals are smoother and more continuous than the ground truth, which explains the resulting lower metric values. These results show that the main advantage provided by CoBSI with $\Gamma_{958}$ in linear sampling is that it can preserve the polarities and location of events on the time axis, pointed by arrows. 

For instance, SBI fails in interpolating the events signaled by arrows in Figure~\ref{fig:E2R}, located close to receiver index 25 and in the time samples 300 and 500, respectively, because it does not have enough neighboring shot information to estimate the correct temporal position of the events. Moreover, CoBSI provides a stronger denoised signal enhancing and highlighting the seismic events in the shot. As in the previous case, the results for Experiment II in Table~\ref{tbl:exps} show that CoBSI-Exp interpolation provides more accurate  results compared to DMSSA, MSSA SGK and SBI, for both metrics. The improvements in this experiment go up to 0.521 (SSIM) and 6.421 dB (PSNR), for the averaged results of all shots.

\begin{figure*}[ht]
    \centering
    \includegraphics[width=\textwidth]{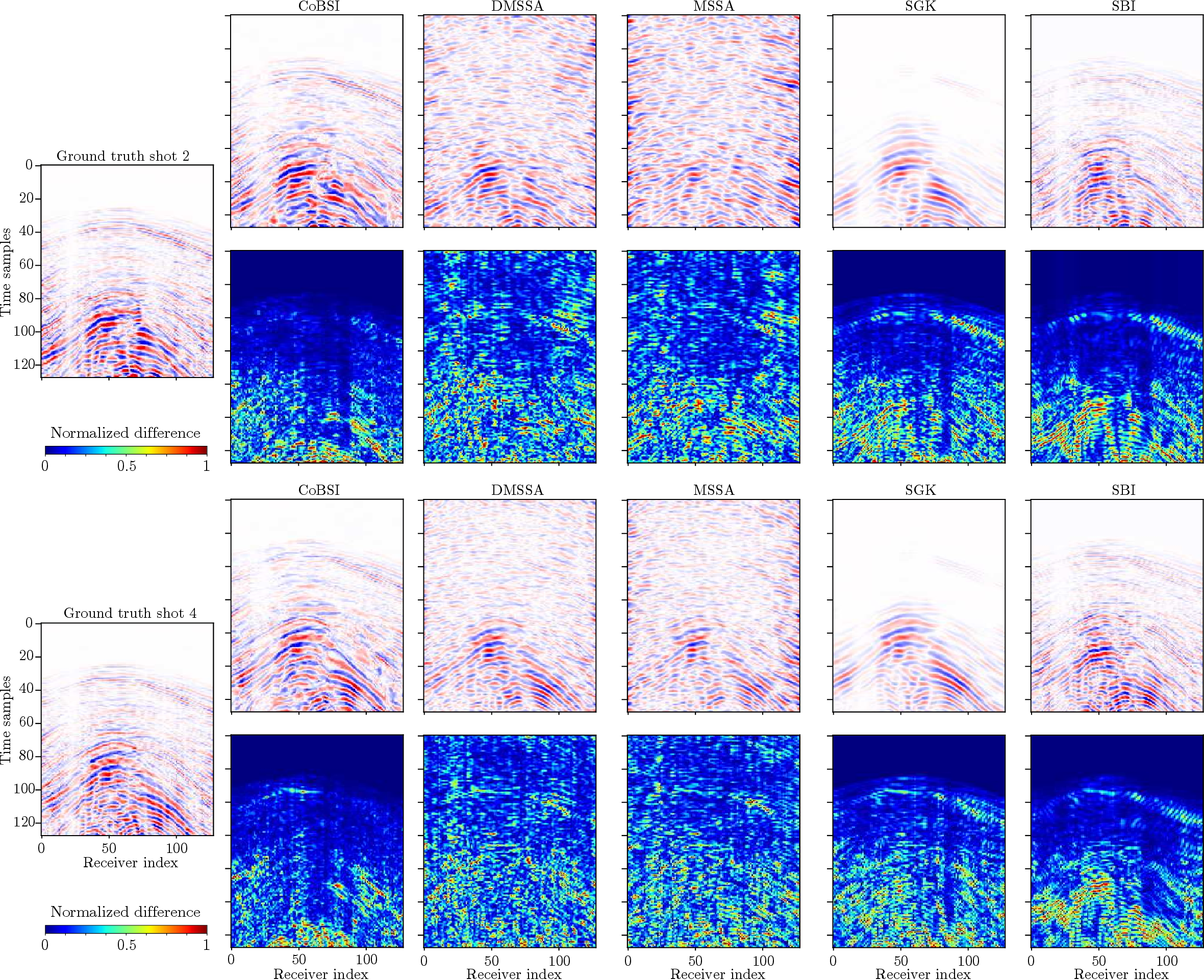}
    \caption{SEAM Phase II data interpolation results for shots S2 and S4 from Experiment III, using CoBSI, DMSSA, MSSA, SGK and SBI methods. Arrows point at events better approximated by CoBSI, while sparsity yields to low amplitude events and artifacts in the main reflection events at the center of the shot.}
    \label{fig:E3r}
\end{figure*}

\textbf{Experiment III:} The data set used in this experiment is a part of the \textit{SEAM Phase II Foothills model}. The acquisition is an orthogonal survey over a complex geological model simulating the Llanos Foothills of the Andes Mountains in Colombia, one of the most challenging regions of active land exploration~\cite{Regone2017}, mainly because of  complex seismic events with abrupt changes in amplitude caused by the topography. The data set comprises  $m=128$ time samples, $n=128$ receivers and $k=7$ shots. The interval samplings are dt=8 ms, dg = 50 m and ds = 50 m for time, receivers and shots, respectively. In this experiment we evaluate the performance of CoBSI in regular acquisitions. Thus, shots S2 and S4 were removed from the survey, and the interpolated shots are illustrated in Figure~\ref{fig:E3r}. It can be seen that CoBSI with $\Gamma_{551}$ in linear sampling reconstructs the events preserving the tilt and polarities, while SBI method yields to events with low amplitude, as well as artifacts in the main reflection events located in the center of the shot. Further, due to the complexity of the distribution of the reflection events, the competing interpolation methods fail in correctly reconstructing the low amplitude signals, this can be seen in the error images, with larger error values  distributed throughout the shot, while in CoBSI the errors occur mostly at the bottom of the shot gather. Therefore, in this scenario CoBSI exhibits higher accuracy as it is able to highlight the main seismic reflection events in the shot gathers. The results for this experiment in Table \ref{tbl:exps} verify this behavior with gains of up to 0.592 (SSIM) and 7.311 dB (PSNR) in the average results.

\section{Discussion}

In general, the previous results demonstrate that CoBSI interpolation outperforms DMSSA, MSSA SGK and SBI for all experiments, providing adequate representations of the typical characteristics of a seismic shot such as smoothness and continuity in the events, noise reduction, and amplitude compensation. The following sub-sections discuss the advantages and limitations of CoBSI related to seismic neural representation and computational costs.

\subsubsection{Seismic Neural Representation} This work aims to show that a complete and continuous representation of seismic data can be obtained by coupling the anisotropic positional encoding and the MLP, despite the simplicity of the neural network, as it has been previously shown for other computational imaging problems \cite{Mildenhall2020, Barron2021}. Further, it should be noted that the CoBSI formulation is flexible, so that other, more complex neural network architectures can be employed instead of the MLP, at the expense of their inherent computational complexity. One of the main advantages of CoBSI is that once the seismic neural representation is obtained from the available incomplete data, it is possible to estimate the response of the continuous field at any arbitrary coordinate within the analysis domain, so that complete missing shot gathers can be accurately estimated. In addition, compared to other interpolation or reconstruction methods where parameter tuning must be done, in CoBSI the number of frequencies in the mapping function is closely related to the dimensionality of the seismic data. Therefore, the hyperparameters $M,N,K$ from Equation \eqref{eq:gammencoding} can be found by analyzing the type of seismic acquisition survey. Specifically, with few seismic sources in a cross-spread array, the signal variation will be mainly focused in the receiver direction (in-line), because more spatial information exists. On the other hand, dealing with inline-offset spread or marine data, where the density of shots is high, the greatest variation of the signal occurs in the source direction (Xline).

\subsubsection{Computational cost} As shown in Table \ref{tbl:summ}, there are fewer than one million trainable parameters in CoBSI. To provide a comparison with respect to more complex network architectures employed in seismic reconstruction, Table \ref{tbl:params} presents the number of trainable parameters (NTP) of convolutional neural networks such as U-Net~\cite{IrregularInterp_DL, Chai2021} and autoencoders~\cite{Mandelli2018, 9751748,Qian2022} under supervised learning schemes, as well as an ordinary convolutional neural network CNN \cite{Wang2020a}. It is worth noting that these methods focus on interpolating receivers in 2D and 3D seismic data. In contrast, CoBSI aims to estimate complete missing shots, which is substantially a more complex task. Further, it is able to obtain such interpolations employing a simpler network (MLP), entailing less computational resources in an unsupervised strategy, as only the weights and bias of the MLP are learned. While~\cite{Wang2020a} is an unsupervised learning approach, it only considers missing trace reconstruction for uniform subsampling, which is unrealistic for field applications.

\begin{table}[ht]
\centering
\caption{Comparison of the number of trainable parameters (NTP) required in seismic reconstruction methods\label{tbl:params}}
\begin{tabular}{llll}
\hline
Seismic Data                                    & Network     & NTP & Learning \\ \hline
2D   \cite{IrregularInterp_DL}         & U-NET       & 87M & Supervised\\ 
2D \cite{Wang2020a}     & CNN ordinary         & 42K & Unsupervised \\ 
3D \cite{Chai2021}     & U-NET       & 27M & Supervised \\
2D \cite{Mandelli2018} & Autoencoder & 18M & Supervised \\ 
3D \cite{Qian2022}     & Autoencoder & 6M & Supervised \\
Shotgather (i.e. 3D)                                &  CoBSI (MLP)        & \textless 1M  & Unsupervised            \\ \hline
\end{tabular}
\end{table}

\section{Conclusions}

A coordinate-based seismic interpolation (CoBSI) method to estimate missing seismic shots in both regular and irregular 3D land seismic acquisitions has been proposed. Unlike state-of-the-art reconstruction methods that employ index-based models, CoBSI employs a coordinate-based approach that allows data interpolation in irregular geometries. Further, a key component of CoBSI is an anisotropic positional encoding layer in the neural network to map from low to high dimension coordinates to consider the variation in the reference axes corresponding to time, receivers and shots domains. Experimental results showed the ability of the proposed method on three different scenarios: (i) irregular geometry and synthetic wavefield, (ii) geometry with a gap in real acquisition from Texas, and (iii) the well-known geophysical \textit{SEAM Phase II Foothills model} with regular acquisition. The obtained results demonstrated the advantages of the proposed method with respect to sparsity-based and low-rank interpolation, since CoBSI can estimate continuous seismic events while providing a smooth signal in the space-time domains. Quantitatively, in average CoBSI outperformed the competing methods by up to 22.11 dB of PSNR in synthetic data and 6.4 dB and 7.31 dB for the real-data experiments.

\section*{Acknowledgment}

This work was supported by the project 110287780575 through the agreement 785-2019 between the \textit{Agencia Nacional de Hidrocarburos} and the \textit{Ministerio de Ciencia, Tecnología e Innovación} and \textit{Fondo Nacional de Financiamiento para la Ciencia, la Tecnología y la Innovación Francisco José de Caldas}, and by the Laboratory Directed Research and Development program of Los Alamos National Laboratory under project number 20200061DR. The authors thank the Centre for High Performance Computing (GUANE-UIS) and NVIDIA Academic Hardware Grant Program for providing computational resources to run the parameter tuning experiments.

\ifCLASSOPTIONcaptionsoff
  \newpage
\fi



\bibliographystyle{IEEEtran}
\bibliography{bib/DOCTORADO}
%

%




\begin{IEEEbiographynophoto}{Paul Goyes-Pen\~afiel}
received the B.Sc. in geology from the Universidad Industrial de Santander, Bucaramanga, Colombia, and M.Sc. in geophysics from the Perm State University, Perm, Russia, in 2009 and 2018 respectively. He currently is a Ph.D. candidate in Computer Science with the Universidad Industrial de Santander. His research interests are in inverse theory and applications in geophysics, seismic acquisition and processing, potential-EM methods and deep learning applications in geoscience.
\end{IEEEbiographynophoto}

\vspace{-0.5in}

\begin{IEEEbiographynophoto}{Edwin Vargas}
(Student Member, IEEE) received the B.Sc. and M.Sc. degrees in electronics engineering from the Universidad Industrial de Santander, Bucaramanga, Colombia, in 2016 and 2018, respectively, where
he is currently pursuing the Ph.D. degree in Electronics Engineering.
His research interests include high-dimensional signal processing, computational imaging, and deep learning.
\end{IEEEbiographynophoto}

\vspace{-0.5in}

\begin{IEEEbiographynophoto}{Claudia V.  Correa}
(Member, IEEE) received the B.Sc. and M.Sc. degrees in computer science from the Universidad Industrial de Santander, Bucaramanga, Colombia, in 2009 and 2013, respectively, and the M.Sc. and Ph.D. degrees in electrical and computer engineering from the University of Delaware, Newark, DE, USA, in 2013 and 2017, respectively. She is currently a research fellow at the Computer Science Department, Universidad Industrial de Santander. Her research interests include  computational imaging, compressive spectral imaging, and machine learning applications to seismic problems.
\end{IEEEbiographynophoto}

\vspace{-0.5in}

\begin{IEEEbiographynophoto}{Yu Sun}
(Student Member, IEEE) received the B.Eng. degree in electronics and information from Sichuan University, Chengdu, China, in 2015, and the M.S. degree in data analytics and statistics in 2017 from Washington University in St. Louis, St. Louis, MO, USA, where he is currently working toward the Ph.D. degree with the Computational Imaging Group. During his Ph.D., he worked as an Intern with Nvidia Corporation in 2021. His research interests include computational imaging, machine learning, deep learning, and optimization.
\end{IEEEbiographynophoto}

\vspace{-0.5in}

\begin{IEEEbiographynophoto}{Ulugbek~S.~Kamilov}
(Senior Member, IEEE)  received the B.Sc./M.Sc.\ degree in communication systems and the Ph.D.\ degree in electrical engineering from EPFL, Lausanne, Switzerland, in 2011 and 2015, respectively. He is an Assistant Professor and the Director of Computational Imaging Group (CIG), Washington University in St.\ Louis, MO, USA. From 2015 to 2017, he was a Research Scientist with Mitsubishi Electric Research Laboratories (MERL), Cambridge, MA, USA. 
Prof.~Kamilov is a Senior Editor of IEEE Signal Processing Magazine, a  Member of Bio Imaging and Signal Processing Technical Committee of the IEEE Signal Processing Society. He was a recipient of the IEEE Signal Processing Society’s 2017 Best Paper Award and the NSF CAREER Award in 2021.
\end{IEEEbiographynophoto}

\vspace{-0.5in}

\begin{IEEEbiographynophoto}{Brendt Wohlberg}
(Fellow, IEEE) received the B.Sc. (Hons.) degree in applied mathematics, and the M.Sc. (applied science) and Ph.D. degrees in electrical engineering from the University of Cape Town, Cape Town, South Africa, in 1990, 1993, and 1996, respectively. He is currently a Staff Scientist with Theoretical Division, Los Alamos National Laboratory, Los Alamos, NM, USA. His primary research interests include signal and image processing inverse problems and computational imaging. He was a co-recipient of the 2020 SIAM Activity Group on
Imaging Science Best Paper Prize. He was an Associate Editor for the IEEE TRANSACTIONS ON IMAGE PROCESSING from 2010 to 2014, and for the IEEE TRANSACTIONS ON COMPUTATIONAL IMAGING from 2015 to 2017, and was the Chair of the Computational Imaging Special Interest Group (now the Computational Imaging Technical Committee) of the IEEE SIGNAL PROCESSING SOCIETY from 2015 to 2017. He was Editor-in-Chief of the IEEE TRANSACTIONS ON COMPUTATIONAL IMAGING from 2018 to 2021, and is currently Editor-in-Chief of the IEEE OPEN JOURNAL OF SIGNAL PROCESSING.
\end{IEEEbiographynophoto}

\vspace{-0.5in}

\begin{IEEEbiographynophoto}{Henry Arguello}
(Senior Member, IEEE) received the B.Sc. Eng. degree in electrical engineering and the M.Sc. degree in electrical power from the Universidad Industrial de Santander, Bucaramanga, Colombia, in 2000 and 2003, respectively, and the Ph.D. degree in electrical engineering from the University of Delaware, Newark, DE, USA, in 2013. He is currently an Associate Professor with the Department of Systems Engineering, Universidad Industrial de Santander and associate editor for IEEE TRANSACTIONS ON COMPUTATIONAL IMAGING. In 2020, he was a Visiting Professor with Stanford University, Stanford, CA, USA, funded by Fulbright. His research interests include high-dimensional signal processing, optical imaging, compressed sensing, hyperspectral imaging, and computational imaging.
\end{IEEEbiographynophoto}





\end{document}